\documentstyle[twocolumn,epsfig]{mn}
\font\japit = cmti10 at 10truept
\include{epsf}

\title
     [Lens magnification by Abell 1689: Background galaxy
luminosity function]
{\vglue-3.0truecm
\centerline{\japit For submission to Monthly Notices}
\vglue 2.5truecm
\noindent
Gravitational lens magnification by Abell 1689:
Distortion of the background galaxy luminosity function}
\author
     [Dye et al.]
     {S. Dye$^1$, A.N. Taylor$^1$, E.M. Thommes$^{1,2}$, K. Meisenheimer$^2$, 
	C. Wolf$^{\,\,2}$, J.A. Peacock$^1$ \\
     $^1$Institute for Astronomy, University of Edinburgh,
     Royal Observatory, Blackford Hill, Edinburgh EH9 3HJ, U.K.\\
     $^2$Max-Planck-Institut f\"{u}r Astronomie, K\"{o}nigstuhl 17,
           D-69117 Heidelberg, Germany}

\def\rmd{{\rm d}}
\def\mpcoh{\,h^{-1}\,{\rm Mpc}}
\newcommand{\be}{\begin{equation}}
\newcommand{\ee}{\end{equation}}
\newcommand{\ba}{\begin{eqnarray}}
\newcommand{\ea}{\end{eqnarray}}

\newcommand{\qvec}{\mbox{\boldmath $q$}}
\newcommand{\Qvec}{\mbox{\boldmath $Q$}}

\newcommand{\Mpc}{{\rm Mpc}}
\newcommand{\lgl}{\langle}
\newcommand{\rgl}{\rangle}
\newcommand{\nn}{\nonumber \\}
\def\bib{\parskip=0pt\par\noindent\hangindent\parindent
    \parskip =2ex plus .5ex minus .1ex}
\begin{document}

\maketitle

\begin{abstract}
Gravitational lensing magnifies the observed flux of galaxies behind
the lens. We use this effect to constrain the total mass in the
cluster Abell 1689 by comparing the lensed luminosities of background
galaxies with the luminosity function of an undistorted field. Under
the assumption that these galaxies are a random sample of luminosity
space, this method is not limited by clustering noise. We use
photometric redshift information to estimate galaxy distance and
intrinsic luminosity. Knowing the redshift distribution of the
background population allows us to lift the mass/background degeneracy
common to lensing analysis. In this paper we use 9 filters observed
over 12 hours with the Calar Alto 3.5m telescope to determine the
redshifts of 1000 galaxies in the field of Abell 1689.  Using a
complete sample of 146 background galaxies we measure the cluster mass
profile. We find that the total projected mass interior to $0.25\mpcoh$
is $M_{2d}(<0.25\mpcoh)=(0.48\pm0.16)\times10^{15}
\,h^{-1}{\rm M}_{\odot}$, where our
error budget includes uncertainties from the photometric redshift
determination, the uncertainty in the offset calibration and finite
sampling. This result is in good agreement with that found by number
count and shear--based methods and provides a new and independent
method to determine cluster masses.

\end{abstract}


\section{Introduction}

The use of gravitational lensing as a means of cluster mass
reconstruction provides a theoretically efficient approach without the
equilibrium and symmetry assumptions which typically accompany virial
and X-ray temperature methods. Mass determination through application
of lens shear proves to give good resolution in mass maps although
measurement of absolute quantities is not possible without external
calibration. This so called sheet-mass degeneracy (Falco, Gorenstein
\& Shapiro 1985) is broken however by methods which exploit the
property of lens magnification.

First recognised by Broadhurst, Taylor \& Peacock (1995, BTP
hereafter) as a viable tool for the reconstruction of cluster mass,
lens magnification has the twofold effect of amplifying background
source galaxy fluxes as well as their geometrical size and
separation. This immediately permits two separate approaches for
measuring lensing mass. The first involves selecting a sample of
sources with a flat or near-flat number count slope. Magnification
results in a reduction of their local surface number density owing to
the dominance of their increased separation over the enhanced number
detectable due to flux amplification.

Although contaminated by faint cluster members, Fort, Mellier \&
Dantel-Fort (1997) first reported this dilution effect using B and I
band observations of the cluster CL0024$+$1654. Later, Taylor et
al. (1998, T98 hereafter) demonstrated how the dilution in surface
number density of a colour-selected sample of red galaxies lying
behind the cluster Abell 1689 enables determination of its total mass
profile and 2d distribution.  A projected mass interior to $0.24\mpcoh$
of $M_{2d}(<0.24\mpcoh)=(0.50\pm0.09)
\times10^{15}\,h^{-1}{\rm M}_{\odot}$ was predicted, in good agreement
with the shear analysis of Tyson \& Fischer (1995) who measured
$M_{2d}(<0.24\mpcoh)=(0.43\pm0.02)\times10^{15}\,h^{-1} {\rm
M}_{\odot}$ and Kaiser (1995) with a measurement of
$M_{2d}(<0.24\mpcoh)=(0.43\pm0.04)\times10^{15}\,h^{-1} {\rm
M}_{\odot}$. 

Since then, several authors have detected source number
count depletion due to cluster lensing. Athreya et al (1999) observe
MS1008$-$1224 and use photometric redshifts to identify a background
population of galaxies within which they measure depletion.  Mayen \&
Soucail (2000) constrain the mass profile of MS1008$-$1224 by comparing
to simulations of depletion curves. Gray et al (2000) measure
the first depletion in the near infra-red due to lensing by Abell
2219. Finally and most recently, R\"{o}gnvaldsson et al. (2000)
find depletion in the source counts behind CL0024$+$1654 in the R band
and for the first time, in the U band.

The second mass reconstruction approach permitted by magnification
forms the primary focus of this paper.  The amplification of flux by
lens magnification introduces a measurable shift in the luminosity
function of background source galaxies. With a sufficiently well
defined luminosity function derived from an unlensed offset field for
comparison, this shift can be measured to allow an estimate of the
lens mass (BTP).  This method relies upon a set of observed source
magnitudes which, if assumed to form an effective random sampling of
luminosity space, is not limited by noise from background source
clustering unlike the number count method (see Section \ref{sec_morph}
for further discussion).

This paper presents the first application of mass reconstruction using
lens flux magnification inferred from the luminosity function of
background samples. Unlike the method of T98 who defined their
background sample based on colour cuts, in this work photometric
redshifts of all objects in the observed field have been
estimated. This not only allows an unambiguous background source
selection but alleviates the need to estimate source distances when
scaling convergence to real lens mass.

The following section details the theory of mass reconstruction from
lens magnification of background source magnitudes. Section
\ref{sec_photo_anal} describes the photometric analysis applied to
observations of A1689 with the redshifts which result. Observations of
the offset field which provide the absolute magnitude distribution
required for comparison with the A1689 background source sample are
presented in Section \ref{sec_cadis_field}.  From this, a
parameterised luminosity function is calculated in Section
\ref{sec_cadis_schechter} necessary for application of the maximum
likelihood method. Following a discussion of sample incompleteness in
Section \ref{sec_completeness}, a mass measurement of A1689 is given
in Section \ref{sec_mass_determ} where the effects of sample
incompleteness are quantified. Finally, a signal to noise study is
carried out in Section \ref{sec_sn_calcs} to investigate the effects
of shot noise, calibration uncertainty of the offset field and
photometric redshift error.

\section{Mass Reconstruction}
\label{sec_mass_recon}

Measurement of lens magnification of background source fluxes
requires a statistical approach in much the same way as do shear
or number count depletion studies. The basis of this statistical
method relies on the comparison of the distribution of lensed source
luminosities with the luminosity function of an un-lensed offset
field. As Section \ref{sec_morph} discusses further, for a fair
comparison, the population of sources detected behind the lens must be
consistent with the population of objects used to form this un-lensed
reference luminosity function.

The effect of lens magnification by a factor $\mu$ on a source is to
translate its observed magnitude from $M$ to
$M+2.5\log_{10}\mu$. In terms of the reference
luminosity function, $\phi(M,z)$, the probability of a background
galaxy with an absolute magnitude $M$ and redshift $z$ being magnified
by a factor $\mu$ is (BTP)
\be
\label{eq_lum_prob}
{\rm P}[M|\mu,z]=\frac{\phi(M+2.5\log_{10}\mu(z),z)} {\int
\phi(M+2.5\log_{10}\mu(z),z)\rmd M}.
\ee 
Magnification depends on the geometry of the observer-lens-source
system hence for a fixed lens and observer, $\mu$ is a function of
source redshift. This redshift dependence comes from the familiar
dimensionless lens surface mass density or
convergence, $\kappa(z)$, and shear, $\gamma(z)$, which are related to 
$\mu(z)$ via,
\be
\label{eq_mag}
\mu(z)=\left|[1-\kappa(z)]^2 - \gamma^2(z)\right|^{-1}.
\ee

We wish to apply maximum likelihood theory using the probability in
equation (\ref{eq_lum_prob}) to determine lens magnification and hence
$\kappa$.  A parametric luminosity function is therefore required and
so we take $\phi(M,z)$ to be a Schechter function (Schechter 1976),
\be
\label{eq_mag_schechter_function}
\phi(M,z)=\phi^*(z)10^{0.4(M_*-M)(1+\alpha)}\exp\left[-10^{0.4(M_*-M)}\right].
\ee 
The Schechter parameters $\phi^*$, $M_*$ and $\alpha$ are determined
by fitting to the magnitude distribution of the offset field (see
Section \ref{sec_cadis_field}).  T98 use $\mu$ as a
likelihood parameter by adopting the simplification that all sources
lie at the same redshift. However this is not possible when each source is
attributed its own redshift. We must therefore express $\mu$ in terms
of a redshift depenent quantity and a source--independent likelihood
parameter. 

The most direct solution is to separate the convergence. Using the
parameter $\kappa_{\infty}$ introduced by BTP as the convergence for
sources at $z=\infty$, we can write $\kappa(z)$ as,
\ba
\label{eq_kappa_zs}
\kappa(z) &=& \kappa_\infty f(z), \nn
f(z) &=& \frac{\sqrt{1+z}-\sqrt{1+z_{\scriptscriptstyle L}}}{\sqrt{1+z}-1}.
\ea
We therefore choose $\kappa_\infty$ as our likelihood parameter with
all source redshift dependency being absorbed into the function $f$.
The lens surface mass density, $\Sigma$, is then related to $\kappa_\infty$
and the lens redshift, $z_{\scriptscriptstyle L}$, by
\ba
\label{eq_kapinf_defn}
\Sigma(z_{\scriptscriptstyle L}) &=& \frac{cH_0}{8 \pi G} \kappa_\infty 
\left[\frac{(1+z_{\scriptscriptstyle L})^2}
{\sqrt{1+z_{\scriptscriptstyle L}}-1}\right] \nn 
&=&  2.75 \times 10^{14} \kappa_\infty\left[
\frac{(1+z_{\scriptscriptstyle L})^2}
{\sqrt{1+z_{\scriptscriptstyle L}}-1}\right] h M_\odot \Mpc^{-2}.
\ea
Here, we assume an Einstein-de-Sitter universe for reasons of simplicity
and because BTP show that this result depends only weakly on the
chosen cosmological model.

Before choosing a likelihood function, consideration must be given
to the shear term in equation (\ref{eq_mag}).  Since shear scales with
source redshift in the same way as the convergence, we use the
so-called $\kappa$ estimators discussed by T98 which relate $\kappa$
to $\gamma$.  At the extremes these are $\gamma=\kappa$ for the
isothermal sphere or $\gamma=0$ for the sheet-like mass. A third
variation, motivated by cluster simulations and the fact that it has
an invertible $\mu(\kappa)$ relation, is $\gamma\propto\kappa^{1/2}$
(van Kampen 1998). This gives rise to the parabolic
estimator which predicts values of $\kappa$ between those
given by the sheet and isothermal estimators. Using equation
(\ref{eq_mag}), the magnification for these three different cases
therefore relates to $\kappa_{\infty}$ via 
\be
\label{eq_mu_redshift_depen}
\mu(z)=\left\{\begin{array}{ll}
\left|1-2\kappa_{\infty}f(z)\right|^{-1} & {\rm iso.} \\
\left|\left[\kappa_{\infty}f(z)-c\right]
\left[\kappa_{\infty} f(z)-1/c\right]\right|^{-1} & {\rm para.} \\
\left[1-\kappa_{\infty}f(z)\right]^{-2} & {\rm sheet}
\end{array}\right. 
\ee 
and hence three different estimations of $\kappa_{\infty}$ exist
for a given $\mu$.  The constant $c$ in the parabolic case is
chosen to provide the best fit with the cluster simulations. As in
T98, we take $c=0.7$ throughout this paper.

The likelihood function for $\kappa_{\infty}$ is then formed from
equation (\ref{eq_lum_prob}),
\be
\label{eq_magnitude_likelihood}
{\cal L}(\kappa_{\infty}) \propto \prod_{i} 
{\rm P}[M_i|\mu(\kappa_{\infty}),z_i]
\ee 
where $\mu$ is one of the three forms in equation
(\ref{eq_mu_redshift_depen}) and the product applies to the
galaxies behind the cluster region under scrutiny.
Absolute surface mass densities are then calculated from $\kappa_{\infty}$
using equation (\ref{eq_kapinf_defn}).

The probability distribution for $\kappa_{\infty}$ obtained from
equation (\ref{eq_lum_prob}) for a single galaxy is typically
double-peaked as two solutions for $\kappa_{\infty}$
exist for a given magnification. The choice of peak is determined by
image parity such that the peak at the higher value of
$\kappa_{\infty}$ is chosen for a galaxy lying inside the critical
line and vice versa. The chosen peak is then extrapolated to extend
over the full $\kappa_{\infty}$ range before contributing
to the likelihood distribution. In this way, a single-peaked likelihood
distribution is obtained.

Evidently, calculation of lens surface mass density in this way
requires redshift and absolute magnitude data for background galaxies
together with knowledge of the intrinsic distribution of magnitudes
from an unlensed offset field. The next section details the
photometric analysis applied to our observations of Abell 1689 to
arrive at background object redshifts and absolute magnitudes.

\section{Photometric Analysis}
\label{sec_photo_anal}

\subsection{Data acquisition}

Observations of Abell 1689 were performed with the Calar Alto 3.5m
telescope in Spain using 8 different filters, chosen for photometric
distinction between foreground, cluster and background objects. In
addition, the I-band observations of T98 were included to bring the
combined exposure time to a total of exactly 12 hours worth of useable
data characterised by a seeing better than $2.1''$. Table
\ref{tab_a1689_filters} details the summed integration time for each
filter set together with the motivation for inclusion of the
filter. Note the narrow band filters 466/8, 480/10 and 774/13 which
were selected to pick out spectral features of objects lying at the
cluster redshift of $z=0.185$ (Teague, Carter \& Gray 1990).

\begin{table}
\vspace{4mm}
\centering
\begin{tabular}{|c|c|c|}
\hline
Filter: $\lambda_c/\Delta\lambda$ (nm) & t$_{\rm int}$(s) & Use \\
\hline
826/137 (I-band) & 6000 & Global SED\\
774/13  & 6800 & H$_\alpha$ at $z=0.185$\\
703/34  & 4100 & Background $z$\\
614/28  & 7700 & Background $z$\\
572/21  & 6300 & Background $z$\\
530/35  & 3300 & Background $z$\\
480/10  & 4200 & 4000\AA at $z=0.185$\\
466/8   & 4800 & Ca H,K at $z=0.185$ \\
457/96 (B-band) & 6000 & Global SED\\
\hline 
\end{tabular}
\caption{The observations of Abell 1689 in all 9 filters (labelled as
$\lambda_c/\Delta\lambda\equiv$ central wavelength/FWHM).
t$_{int}$ gives the total integration time in each filter. The
I-band data comes from T98.}
\label{tab_a1689_filters}
\end{table}

Image reduction and photometry was performed using the MPIAPHOT
(Meisenheimer \& R\"{o}ser 1996) software written at the MPIA
Heidelberg as an extension to the MIDAS reduction software package.
Images were de-biased and flattened from typically four or five
median filtered dusk sky flats observed each night for each filter
set. Any large scale remnant flux gradients were subsequently removed
by flattening with a second-order polynomial fitted to the image
background. Cosmic ray removal was carried out using the pixel
rejection algorithm incorporated in MPIAPHOT. All post-reduced images
were flattened to a $1\sigma$ background flux variation of less than
0.02 mag.

\subsection{Galaxy catalogue}
\label{sec_gal_cat}

Instead of co-adding images in each filter set before object
detection, photometric evaluation was carried out on images
individually. In this way, an estimate of the uncertainty in the photon
count for each galaxy could be obtained.
The mean photon count $I^{(b,m)}$
of a galaxy $m$ observed in a filter $b$ was calculated as the usual
reciprocal-variance weighted sum, 
\be
\label{eq_flux_weight}
I^{(b,m)}=
\sum_i\frac{I^{(b,m)}_i}{\left(\sigma_i^{(b,m)}\right)^2}
{\Big /}\left[\sum_i\left(\sigma_i^{(b,m)}\right)^{-2}\right]
\ee
where the summation acts over all images belonging to a particular
filter set and the error on $I^{(b,m)}$ is
\be
\label{eq_flux_weight_error}
\overline{\sigma}^{(b,m)}=\left[\sum_i\left(
\sigma_i^{(b,m)}\right)^{-2}\right]^{-1/2} .
\ee
The quantity $\sigma_i^{(b,m)}$ is the standard deviation of
background pixel values surrounding galaxy $m$ in image
$i$. Background pixels were segregated by applying an appropriate cut
to the histogram of counts in pixels within a box of size $13''\times
13''$ ($40\times 40$ pixels) centred on the galaxy. This cut removed
the high count pixels belonging to the galaxy itself and any other
neighbouring galaxies within the box.

Integrated galaxy photon counts were determined using MPIAPHOT which
sums together counts in all pixels lying inside a fixed aperture of
radius $6''$ centred on each galaxy. A `mark table' accompanying every
image in every filterset provided co-ordinates of galaxy centres.
Tables were fit within an accuracy of $<1''$ to individual images
using copies of a master table derived from the deepest co-added
image; that observed in the I-band.  In this way consistent indexing
of each galaxy was achieved throughout all catalogues. The master
table was generated using the object detection software `SExtractor'
(Bertin \& Arnouts 1996). Only galaxies were contained in the master
table, star-like objects being removed after identification by their
high brightness and low FWHM.  With a detection threshold of $3\sigma$
above the average background flux, galaxies in the I-band image were
catalogued after coincidence matching with objects detected at the
$3\sigma$ level in the associated V-band data presented by
T98. Despite cataloguing $\sim 3000$ galaxies, the resulting number
was limited to a total of $\sim 1000$ due to the relatively shallow
data observed with the 466/8 narrow-band filter.

\subsection{Photometry}
\label{sec_photom}

The integration of photon counts in an aperture of fixed size requires
constant seeing across all images to allow correct determination of
colours.  To ensure constant seeing, all images were degraded by
Gaussian convolution to the worst seeing of $2.1''$ measured in the
466/8 filter before galaxy counts were evaluated. The effects of
changing weather conditions were compensated for by normalising images
within each filter set to an arbitrarily chosen image in that
set. Normalisation was conducted by scaling the galaxy counts in each
image so that the average counts of the same stars in all images was
equal. This ensured correct calculation of the weighted counts and the
error from equation (\ref{eq_flux_weight}) and
(\ref{eq_flux_weight_error}). These quantities were later scaled to
their calibrated photometric values.

Calibration of the photometric fluxes from the weighted counts
was provided using the spectrum of the dominant elliptical galaxy in
the centre of A1689 taken from Pickles \& van der Kruit
(1991). Denoting this spectrum as the function $F_s(\lambda)$, the
calibration scale factors $k_b$ for each filter $b$ were calculated
using
\be
\label{eq_standard_flux}
I^{(b,s)}=k_b\int \rmd\lambda
\frac{E(\lambda)T_b(\lambda)F_s(\lambda)\lambda}{hc}
\ee 
where the function $T_b(\lambda)$ describes the filter transmission
efficiency, $E(\lambda)$ is the combined filter-independent efficiency
of the detector and telescope optics and $I^{(b,s)}$ is the measured
integrated photon count rate of the central galaxy. The values of
$k_b$ obtained in this way were only relatively correct owing to the
lack of an absolute calibration of the published spectrum. Absolute
calibration scale factors were calculated from observations of the
standard star G60-54 (Oke 1990) in all filters in exactly the same
manner. Verification of this absolute calibration was provided by the
consistency of ratios of $k_b({\rm absolute})/k_b({\rm relative})$ to a
zero point of $\Delta m=2.11\pm 0.01$ magnitudes averaged over all
filters.

Consideration of equation (\ref{eq_standard_flux}) shows that only the
quantity 
\be
\label{eq_photom_integral}
\int \rmd\lambda E(\lambda)T_b(\lambda)F_m(\lambda)\lambda = 
\frac{hc \, I^{(b,m)}}{k_b}
\ee
can be known for any galaxy $m$ with a calibrated photon count rate.
The required photometric flux 
\be
\label{eq_filter_inten}
F^{(b,m)}=\int\rmd\lambda T_b(\lambda)F_m(\lambda)
\ee
can not therefore be directly determined
without making an approximation such as the simplification of
filter transmission curves to top hat functions. Although this is
acceptably accurate in narrow band filters, it is not
for broad band filters. 
This problem was avoided by the more sophisticated technique of
fitting model spectra to measured galaxy colours 
as the next section discusses.

\subsection{Photometric redshift evaluation}

Direct calculation of photometric fluxes using equation
(\ref{eq_filter_inten}) was made possible by fitting model spectra to
the set of calibrated photon count rates measured for each galaxy
across all filters. Expressed more quantitatively, equation
(\ref{eq_photom_integral}) was applied for each filter to a library of
template spectra to arrive at a set of scaled filter counts for each
spectrum. Galaxies were then allocated library spectra by finding the
set of library colours which best fit the measured galaxy colours.
Note that this differs from conventional template fitting where 
spectra are redshifted and scaled to fit observed colours in a much
more time costly manner.

The spectral library was formed from the template galaxy spectra of
Kinney et al. (1996). A regular grid of galaxy templates was
generated, varying in redshift along one axis from $z=0$ to $z=1.6$ in
steps of $\Delta z=0.002$ and ranging over 100 spectral types from
ellipticals, through spirals to starbursts along the other. 

The set of photometric errors given by equation
(\ref{eq_flux_weight_error}) for an individual galaxy across all
filters gives rise to an error ellipsoid in colour space.  Using the
size and location of these error ellipsoids, probabilities of each
library entry causing the observed sets of colours for each galaxy
were then calculated as 
\be
\label{eq_prob_colour_fit}
p(\qvec|z,s)=\frac{1}{\sqrt{(2\pi)^n|V|}}
\exp\left(-\frac{1}{2}\sum_{j=1}^n\frac{[q_j-Q_j(z,s)]^2}{\sigma_j^2}\right)
\ee 
where $n$ is the number of colours, $\sigma_j$ comes from propagation
of the error given by equation (\ref{eq_flux_weight_error}) and
$V\equiv{\rm diag}(\sigma_1^2,...,\sigma_n^2)$.  Each galaxy's position
vector in colour space, $\qvec\equiv(q_1,...,q_n)$ is compared with
the colour vector $\Qvec$ of the library spectrum with a given
redshift $z$ and type $s$. Finding the maximum probability
corresponding to the closest set of matching colours therefore
immediately establishes redshift and galaxy type. An assessment of the
uncertainty in this redshift is subsequently obtained directly from
the distribution of the probabilities associated with neighbouring
library spectra.
 
\begin{figure}
\vspace{4mm}
\epsfxsize=82mm
{\hfill
\epsfbox{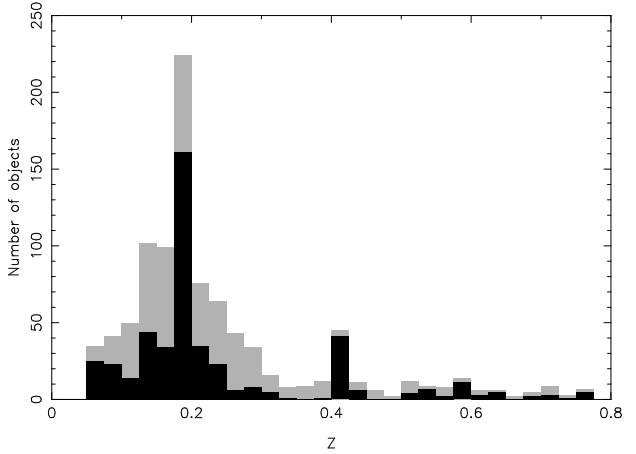}
\hfill}
\epsfverbosetrue
\caption{\small Redshift distribution of the 958 objects photometrically
evaluated in the field of A1689. The grey histogram plots all 958
redshifts whereas the black histogram plots only the 470 redshifts
with a $1\sigma$ error in redshift of less than 0.05. The peak at
$z\simeq0.18$ is the contribution from the cluster galaxies.}
\label{a1689_z_dist}
\end{figure}

Figure \ref{a1689_z_dist} shows the distribution of the 958
successfully classified galaxy redshifts estimated from the full
filter set.  We measure an average redshift error of 
$\left<\sigma_z\right>=0.08$.
The maximum redshift limit of $z\simeq 0.8$ comes from
the condition that the 4000\AA\ limit must lie in or blue-ward of the
second reddest filter in the set.  The peak at $z\simeq 0.18$ in
Figure \ref{a1689_z_dist} is clearly the contribution from the cluster
galaxies. 

The feature at $z\simeq 0.4$ is most likely real and not an artifact
of the photometric method. Such artifacts occur due to `redshift
focusing' when particular redshifts are measured more accurately than
others. Where the uncertainty is larger, galaxies can be randomly
scattered out of redshift bins, producing under-densities and
corresponding over-densities where the redshift measurement is more
accurate. This effect depends on the details of the filter set, being
more common when fewer filters are used, but can be modelled by Monte
Carlo methods.

\begin{figure}
\vspace{4mm}
\epsfxsize=70mm
{\hfill
\epsfbox{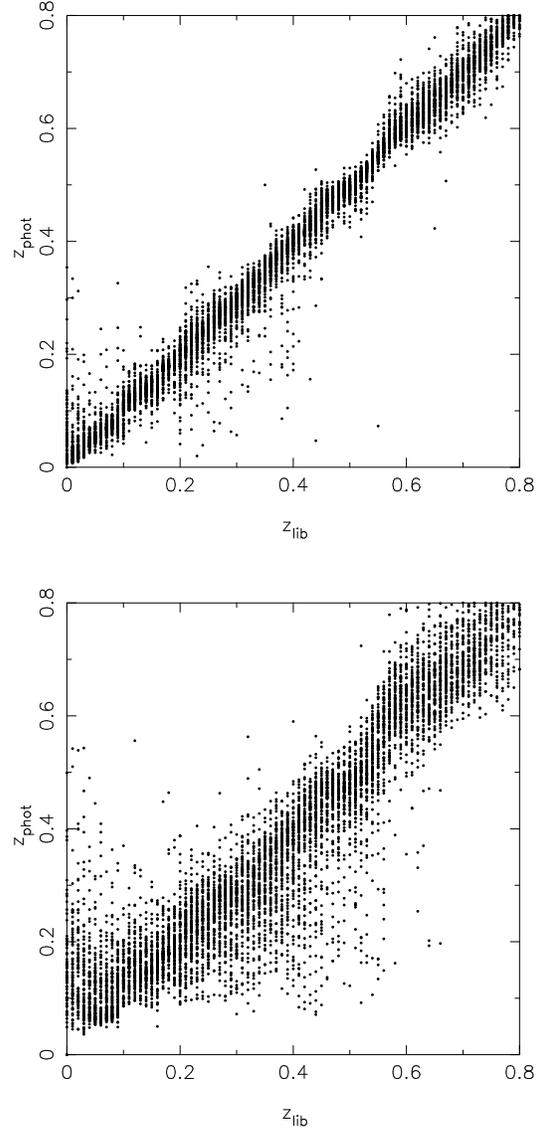}
\hfill}
\epsfverbosetrue
\caption{\small A single Monte Carlo realisation showing the accuracy
of the photometric redshift evaluation method.  Input library spectra
with redshifts $z_{\rm lib}$ are scaled to ${\rm I}=20$ (top) and
${\rm I}=21$ (bottom) and subsequently used to calculate sets of
colours using the A1689 filterset. These colours are randomly
scattered by the filter-specific photometric errors measured in the
A1689 data before calculating the reproduced redshifts $z_{\rm phot}$.}
\label{z_match}
\end{figure}

The top half of Figure \ref{z_match} shows the results of one
realisation of such a Monte Carlo test for redshift focusing. The plot
indicates how accurately the method reproduces redshifts of spectra
scaled to ${\rm I}=20$ with photometric noise levels
taken from the A1689 filter set. Each point represents a single
library spectrum.  Reproduced spectral redshifts, $z_{\rm phot}$, were
determined by calculating colours through application of equation
(\ref{eq_filter_inten}) to the library spectra with redshifts $z_{\rm
lib}$. These colours were then randomly scattered by an amount
determined from the filter-specific photometric error measured in the
A1689 data before application of the redshift estimation method
outlined above. The bottom half of Figure \ref{z_match} shows the same
plot generated using spectra scaled to ${\rm I}=21$ with the same
photometric error taken from the A1689 data.

The accuracy of reproduced redshifts at ${\rm I}=20$ is clearly better
than those at ${\rm I}=21$ where photometric noise is more
dominant. The lack of any sign of redshift focusing in the vicinity of
$z\simeq 0.4$ leads us to conclude that the feature seen at this
redshift in the A1689 data is probably real. The ${\rm I}=21$ plot
which corresponds approximately to our sample magnitude cut of ${\rm
B}=23.6$ (see Section \ref{sec_completeness}) shows that galaxies at
redshifs below $z=0.05$ have on average higher estimated redshifts.
This only marginally affects the overall redshift distribution and yet
partly explains the lack of galaxies at $z<0.05$ in the A1689
redshifts of Figure \ref{a1689_z_dist}. It is worth emphasising here
that the significance of the peak at $z\simeq0.18$ attributed to the
cluster galaxies is far in excess of any effects of redshift focusing.

\begin{figure}
\vspace{4mm}
\epsfxsize=70mm
{\hfill
\epsfbox{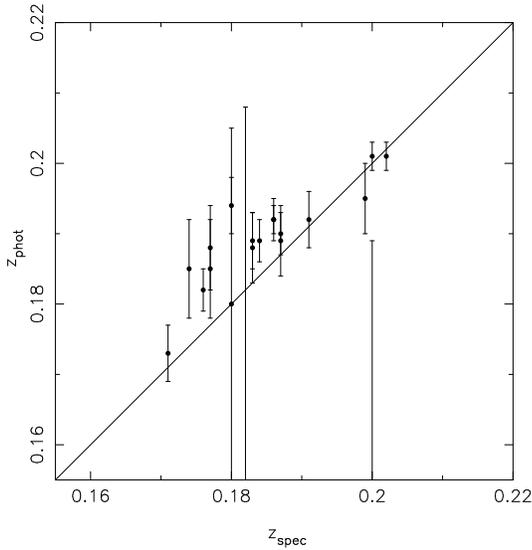}
\hfill}
\epsfverbosetrue
\caption{\small Comparison of the photometric redshifts estimated in the 
cluster Abell 1689 with spectroscopically determined redshifts. The 
distribution shows slight non-Gaussianity in the error distribution. The
mean redshift of the cluster determined spectroscopically is $z=0.185$
(Teague, Carter \& Gray 1990), while the mean photometric redshift is $z=0.189
\pm 0.005$.}
\label{photo_reliability}
\end{figure}

Figure \ref{photo_reliability} shows a comparison of the
photometrically determined redshifts $z_{\rm phot}$ around the peak of
the redshift distribution of Abell 1689, with spectroscopically
determined (Teague, Carter \& Gray, 1990) redshifts $z_{\rm spec}$. A
very slight bias between $z_{\rm phot}$ and $z_{\rm spec}$ can be
seen. This bias is quantified by fitting the line $z_{\rm phot}=z_{\rm
spec}+c$ by least-squares to the data points which gives
$c=0.0036$. Referring to equation (\ref{eq_error_fits}) shows that if
this small bias is applied to all redshifts in our sample, a
negligible difference of $\Delta\kappa_\infty=0.001$ would result. Our
filter set was selected primarily to distinguish the cluster members,
hence at higher redshift we must rely on our Monte Carlo estimates of
the redshift uncertainty (see Section \ref{sec_source_uncert}).

Abell 1689 lies in a region of sky where there is a very low level
of galactic dust. Our redshifts are therefore not affected by this
source of contamination. However, dust in the cluster itself
is another concern. We have modelled the effects of reddening by
cluster dust and find that although magnitudes are slightly affected,
the redshifts remain the same.

\section{Offset field and lens calibration}
\label{sec_cadis_field}

The unlensed, intrinsic magnitude distribution required
by the likelihood analysis was taken from an offset field observed as
part of the Calar Alto Deep Imaging Survey (CADIS) conducted by the
Max-Planck Institut f\"{u}r Astronomie, Heidelberg (Meisenheimer et
al. 1998). Data for this survey were observed to a complete depth of
${\rm B}\simeq24.5$ mag in 16 filters from the B-band to the K-band
with the 2.2m telescope at Calar Alto. We use the CADIS 16-hour field
for our mass calibration.

Using exactly the same methods outlined in the previous section for
the A1689 data, photometric redshifts and rest-frame absolute
magnitudes were determined for all objects in the field.  In addition
to galaxy templates in the spectral library however, 
quasar spectra from Francis et al. (1991) and stellar spectra from
Gunn \& Stryker (1983) were also included, the primary motivation for
this difference being the CADIS quasar study. As a by-product, a more
sophisticated object classification method was achieved by finding the
overall object class yielding the highest significant probability
given by equation (\ref{eq_prob_colour_fit}). Details of this and the
CADIS quasar study are given in Wolf et al. (1999).  To ensure a fair
comparison between the offset field and the cluster field, the CADIS
B-band ($\lambda_c/\Delta\lambda=461/113$ nm) galaxy magnitudes were
used with the A1689 B-band magnitudes in the likelihood analysis 
discussed in Section \ref{sec_mass_recon}.

Investigation of evolution (see for example, Lilly et al. 1995, Ellis
et al. 1996) of the CADIS luminosity function is left for future
work. A preliminary study indicated no significant evolution which
would impact on the lens mass determination. We therefore applied the
same redshift selection as used for the Abell 1689 background sources,
and assumed a no--evolution model.

We present two estimates of the calibration B-band luminosity function:
a nonparametric $1/V_{\rm max}$ method (Section \ref{sec_vmax}) and a  
maximum likelihood parametric fit to a Schechter function 
(Section \ref{sec_cadis_schechter}). The former method allows us to
see the distribution of luminosities without imposing a
preconceived function, and gives a visual impression of the uncertainties
in the parametric fit. In addition the $V_{\rm max}$ approach allows 
us to make basic tests for sample completeness (we discuss this in 
Section \ref{sec_completeness}). The latter maximum likelihood fit provides
a convenient function for performing the second likelihood analysis to
determine lens magnification. We begin by describing the 
nonparametric method.

\subsection{The nonparametric CADIS B-band luminosity function}
\label{sec_vmax}

An estimate of the luminosity function of galaxies in the CADIS B band
was provided initially using the canonical $1/V_{\rm max}$ method
of Schmidt (1968). The
quantity $V_{\rm max}$ is computed for each galaxy as the comoving volume
within which the galaxy could lie and still remain in the redshift and
magnitude limits of the survey. For an Einstein-de-Sitter universe,
this volume is,
\be
\label{eq_vmax}
V_{\rm max}=\left(\frac{c}{H_0}\right)^3\delta\omega
\int^{{\rm min}(z_u,z_{m_{\rm max}})}_{{\rm max}(z_l,
z_{m_{\rm min}})}\rmd z \frac{D^2(z)}{(1+z)^{3/2}} \nn
\ee
where $\delta\omega$ is the solid angle of the observed field of view
and $D(z)$ is
\be
D(z)=2(1-(1+z)^{-1/2}).
\ee
The upper limit of the integral in equation (\ref{eq_vmax}) is set by
the minimum of the upper limit of the redshift interval chosen, $z_u$,
and the redshift at which the galaxy would have to lie to have an
apparent magnitude of the faint limit of the survey,
$z_{m_{\rm max}}$. Similarly, the maximum of the lower limit of the chosen
redshift interval, $z_l$, 
and the redshift at which the galaxy would
have to lie to have an apparent magnitude of the bright limit of the
survey, $z_{m_{\rm min}}$, forms the lower limit of the integral. This
lower integral limit plays a non-crucial role when integrating over
large volumes originating close to the observer where the volume
element makes only a relatively small contribution to $V_{\rm max}$.

The redshifts $z_{m_{\rm max}}$ and $z_{m_{\rm min}}$ are calculated for each
object by finding the roots of
\be
\label{eq_app_abs_mag}
M-m_{\rm lim}+5\log_{10}\left[\frac{(1+z)D(z)}{h_0}\right] -K(z)=-42.39 
\ee 
where $M$ is the absolute magnitude of the object, $m_{\rm lim}$ is the
appropriate maximum or minimum survey limit and $K(z)$ is the
K-correction.  Although the K-correction for each object at its actual
redshift was known from its apparent and absolute rest-frame
magnitude, the redshift dependence of this K-correction was not. In
principle, this redshift dependence could have been calculated
directly for each object using its best-fit spectrum returned from the
photometric analysis. However, the much simplified
approach of approximating the K-correction as a linear function of
redshift was employed. This was primarily motivated by its
improved efficiency and the relatively weak influence the K-correction
was found to have on the final luminosity function.
Lilly et al. (1995) find that the K-correction in their B-band for elliptical, 
spiral and irregular galaxies is proportional to redshift in the redshift 
range $0\leq z \leq 0.8$. The following form for $K(z)$ 
was thus adopted 
\be
\label{eq_kcorr_approx}
K(z)=\frac{K(z_0)}{z_0}z
\ee
where $K(z_0)$ is the K-correction of the object at its actual
redshift $z_0$. 

Once $V_{\rm max}$ has been calculated for all objects, the 
maximum likelihood estimated luminosity function
$\phi$ at the rest-frame absolute magnitude $M$ in bins of width
$\rmd M$ is then,
\be
\label{eq_vmax_lumfn}
\phi(M) \rmd M = \sum_i \frac{1}{V_{{\rm max},i}}
\ee
where the sum acts over all objects with magnitudes between $M-\rmd
M/2$ and $M+\rmd M/2$. 

\begin{figure}
\vspace{4mm}
\epsfxsize=83mm
{\hfill
\epsfbox{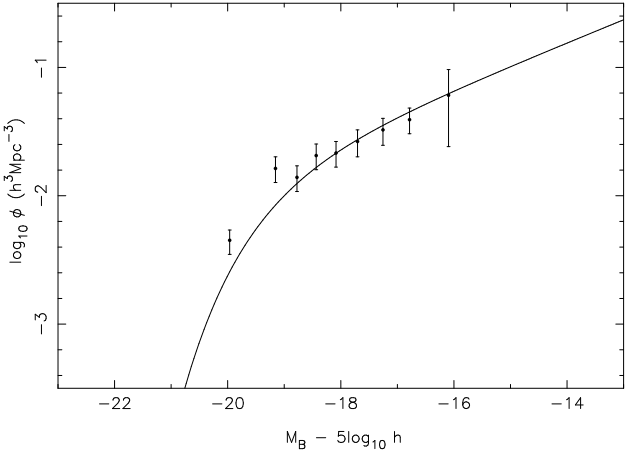}
\hfill}
\epsfverbosetrue
\caption{\small The CADIS B band object luminosity function calculated with
the $1/V_{\rm max}$ formalism. Errors account only for errors in
redshift. Points lie at bin centres, the widths of each chosen to hold
the same number of objects. There are 371 galaxies in total selected by
the redshift limits $0.3\leq z \leq 0.8$ and the apparent B magnitude
$m_{\rm B} \leq 24.5$. The solid line is the Schechter function determined
in Section \ref{sec_cadis_schechter}.}
\label{cadis_lf_vmax}
\end{figure}

Figure \ref{cadis_lf_vmax} shows the luminosity function of B band
magnitudes from the CADIS offset field which has a solid viewing angle
of $\delta \omega =100 \,\, \rm{arcmin}^2$. To match the selection of
objects lying behind A1689, only objects within the redshift range
$0.3\leq z \leq 0.8$ were chosen. A further restriction on the
apparent B magnitude of $m_{\rm B} \leq 24.5$ was applied for completeness
of the sample (see Section \ref{sec_completeness}), yielding a
total of 371 galaxies. The data points in Figure \ref{cadis_lf_vmax}
are centred on bins chosen to maintain an equal number of objects in
each. 

The $1\sigma$ errors shown here were calculated from
Monte Carlo simulations. Object redshifts were randomly scattered
in accordance with their associated errors provided in the
CADIS dataset. For each realisation, the $V_{\rm max}$ of each
object was re-calculated using the re-sampled redshift. The resulting
standard deviation of the distribution of values of $\phi$ for each
bin given by equation (\ref{eq_vmax_lumfn}) was then taken as the
error.  In this particular instance, no consideration was given to the
magnitude errors or the propagation of the redshift error into object
magnitudes. Section \ref{sec_cadis_schechter} discusses this
further. 

\subsection{Parameterisation of the CADIS B-band luminosity function}
\label{sec_cadis_schechter}

The maximum likelihood method of Sandage, Tammann \& Yahil (1979, STY
hereafter) was employed to determine the Schechter function best
describing the CADIS B-band magnitudes. This parameterisation is
essential for the determination of lens mass using the likelihood
approach.

In much the same way as the probability in equation
(\ref{eq_lum_prob}) was formed, the STY method forms the probability
$p_i$ that a galaxy $i$ has an absolute magnitude $M_i$,
\be
\label{eq_schechter_p_i}
p_i\equiv p(M_i|z_i)\propto\frac{\phi(M_i)}
{\int^{{\rm min}(M_{\rm max}(z_i),M_2)}_{{\rm max}
(M_{\rm min}(z_i),M_1)}\phi(M)\rmd M }
\ee
where $M_{\rm max}(z_i)$ and $M_{\rm min}(z_i)$ are the absolute magnitude
limits corresponding to the apparent magnitude limits of the survey at
a redshift of $z_i$. Conversion of these apparent magnitude limits
includes the K-correction using equation (\ref{eq_app_abs_mag})
with $z$ set to $z_i$. A further restriction is placed upon the
integration range by imposing another set of magnitude limits
$M_1<M<M_2$ which for the CADIS data were set at the maximum and
minimum absolute magnitudes found in the sample.

The likelihood distribution in this case is a two dimensional function
of the Schechter parameters $M_*$ and $\alpha$ formed from the product
of all probabilities $p_i$. The best fit $M_*$ and $\alpha$ are
therefore found by maximizing the likelihood function, 
\be
\label{eq_schechter_likelihood}
\begin{array}{l} 
\ln {\cal L}(M_*,\alpha)= \\
\sum_{i=1}^{N}\left\{\ln\phi(M_i)-\ln\int^{{\rm min}(M_{\rm max}
(z_i),M_2)}_{{\rm max}(M_{\rm min}(z_i),M_1)}\phi\,\rmd M \right\} + c_p
\end{array}
\ee
with the constant $c_p$ arising from the proportionality in equation
(\ref{eq_schechter_p_i}). An estimate of the errors on $M_*$ and
$\alpha$ are calculated by finding the contour in $\alpha,M_*$ space
which encompasses values of $\alpha$ and $M_*$ lying within a
desired confidence level about the maximum likelihood
${\cal L}_{\rm max}$. 

To account for uncertainties in the Schechter parameters due to the
redshift and magnitude errors derived by the photometric analysis,
Monte Carlo simulations were once again performed. Redshifts and
absolute magnitudes of the entire sample were randomly scattered
before re-application of the maximum likelihood process each time. The
error in absolute magnitude was calculated using simple error
propagation through equation (\ref{eq_app_abs_mag}) yielding,
\be
\label{eq_z_mag_error_prop}
\sigma^2_{M_i}=\left(\frac{K_i}{z_i}-\frac{5}{\ln 10}\,
\frac{1-0.5(1+z_i)^{-1/2}}{1+z_i-(1+z_i)^{1/2}}\right)^2
\sigma^2_{z_i}+\sigma^2_{m_i}
\ee
for each object with redshift $z_i$ and apparent magnitude
$m_i$. Here, the K-correction given by equation
(\ref{eq_kcorr_approx}) has been used such that the quantity $K_i
\equiv K(z_i)$ is calculated from equation (\ref{eq_app_abs_mag})
using $m_i$, $M_i$ and $z_i$ as they appeared in the CADIS dataset.
The typical ratio of apparent magnitude to redshift error was found to
be $\sigma_{m_i}/\sigma_{z_i}\sim 1\%$ due to the relatively imprecise
nature of photometric redshift determination.

\begin{figure}
\vspace{4mm}
\epsfxsize=80mm
{\hfill
\epsfbox{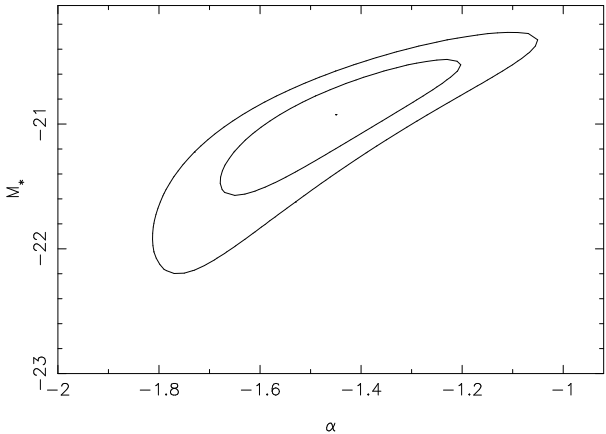}
\hfill}
\epsfverbosetrue
\caption{\small Likelihood contours for the CADIS B-band Schechter
parameters taking photometric redshift and magnitude error into
consideration. $1\sigma$ and $2\sigma$ confidence levels are plotted
corresponding to $\Delta \ln {\cal L}=1.15$ and $3.09$ respectively.}
\label{cadis_likelihood}
\end{figure}

The final errors on $M_*$ and $\alpha$ were taken from an effectively
convolved likelihood distribution obtained by combining the scattered
distributions produced from the Monte Carlo simulations. Figure
\ref{cadis_likelihood} shows the final $1\sigma$ and $2\sigma$
likelihood contours calculated allowing for redshift and magnitude
errors. These predict the resulting parameters
\be
\label{params_ms_alpha_cadis}
M_*=-19.43^{+0.47}_{-0.64} +5\log_{10}h 
\, , \quad \alpha=-1.45^{+0.25}_{-0.23}
\ee
where projected $1\sigma$ errors are quoted. For completeness, 
the normalisation was also calculated to be
\be
\phi^*=0.0164\,h^3 \, {\rm Mpc}^{-3},
\ee
in this case where evolution of the luminosity function has been
neglected.

\section{Sample Consistency}

Ensuring that the A1689 sample of sources is consistent with the CADIS
offset field sample is necessary to prevent biases from entering our
results. The first level of compatibility we have already enforced by
applying a redshift selection of $0.3\leq z \leq 0.8$ to both samples.
The second, discussed in Section \ref{sec_completeness} below, is
sample completeness. A slightly less obvious consideration must also
be given to galaxy morphological type as Section \ref{sec_morph}
explains.

\subsection{Completeness}
\label{sec_completeness}

Determination of the faint magnitude limit beyond which both the A1689
and CADIS data set become incomplete is important for the calculation
of an accurate lens mass. Both samples must be complete for fair
comparison.  Incorrect evaluation of the CADIS limiting magnitude
results in larger values of $V_{\rm max}$ and hence a biased luminosity
function not representative of the intrinsic A1689 distribution. Similarly, 
completeness of the A1689 sample also affects the lens mass
in a manner quantified in Section \ref{sec_incompleteness_effects}.

Estimation of the completeness of both data sets was provided using
the $V/V_{\rm max}$ statistic (Schmidt 1968).  In this ratio, $V_{\rm
max}$ is calculated using equation (\ref{eq_vmax}) whereas $V$ is the
comoving volume described by the observer's field of view from the
same lower redshift limit in the integral of equation (\ref{eq_vmax})
to the redshift of the object. If a sample of objects is unclustered,
exhibits no evolution and is complete, the position of each object in
its associated volume $V_{\rm max}$ will be completely random. In this
case, the distribution of the $V/V_{\rm max}$ statistic over the range 0
to 1 will be uniform with $\langle V/V_{\rm max}\rangle=0.5$.

If the sample is affected by evolution such that more intrinsically
bright objects lie at the outer edges of the $V_{\rm max}$ volume,
then $V/V_{\rm max}$ is biased towards values larger than 0.5. The
reverse is true if a larger number of brighter objects lie nearby. If
the sample is incomplete at the limiting apparent magnitude chosen,
estimations of $V_{\rm max}$ will be on average too large and will
cause $V/V_{\rm max}$ to be biased towards values less than 0.5. The
requirement that $\langle V/V_{\rm max}\rangle=0.5$ for completeness is also
subject to fluctuations due to finite numbers of objects. In the
absence of clustering, the uncertainty due to shot noise on
$\langle V/V_{\rm max}\rangle$ calculated from $N$ galaxies can 
be simply shown to be
\be
\label{eq_sigma_v_vmax_av}
\sigma^2_{\langle V/V_{\rm max}\rangle}=\frac{1}{12N}.
\ee

\begin{figure}
\vspace{4mm}
\epsfxsize=80mm
{\hfill
\epsfbox{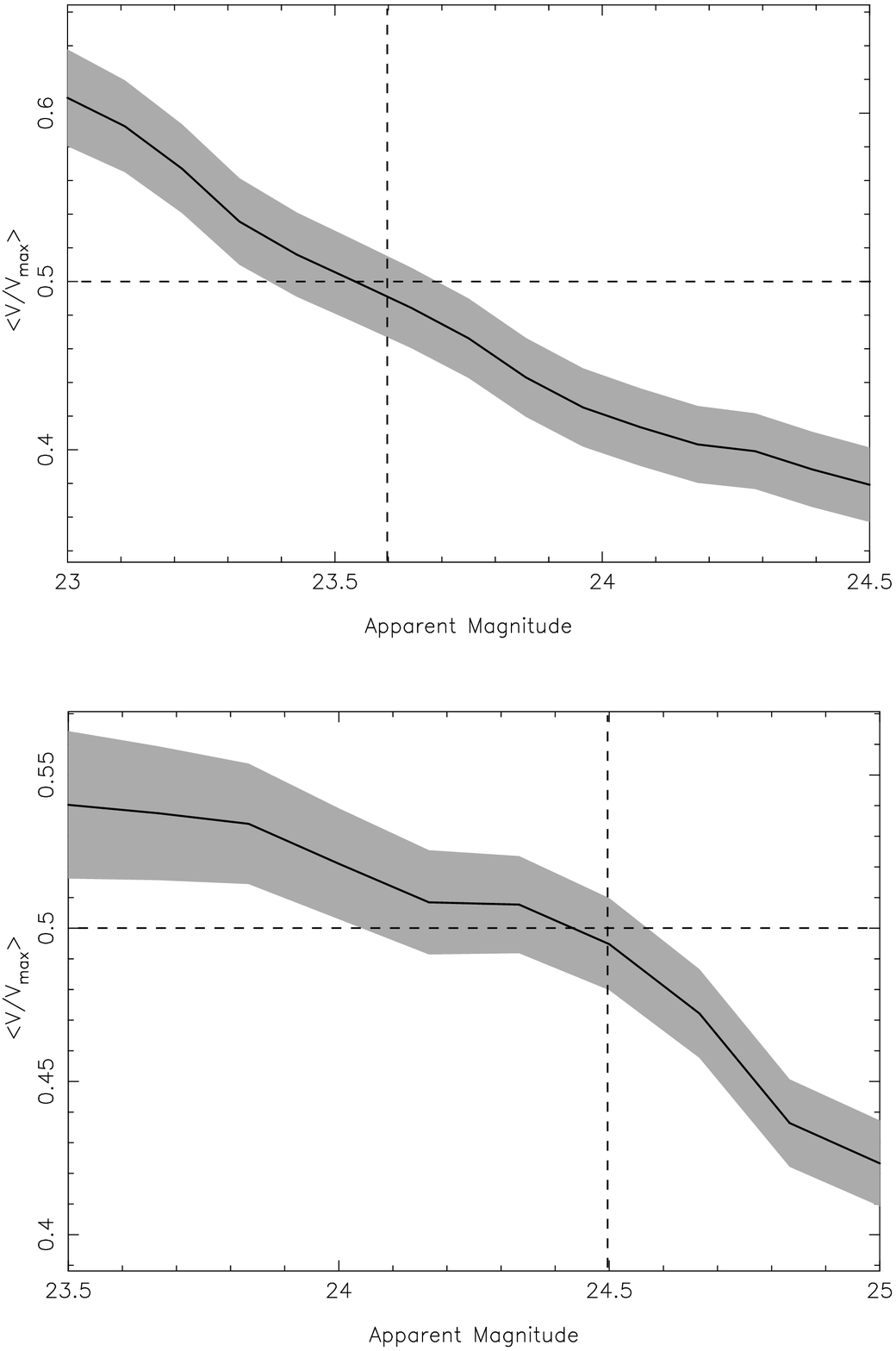}
\hfill}
\epsfverbosetrue
\caption{\small Variation of $V/V_{\rm max}$ with limiting apparent B
magnitude for the A1689 (top) and the CADIS (bottom) sample. The grey
region corresponds to the $1\sigma$ errors described by equation
(\ref{eq_sigma_v_vmax_av}) which are an underestimate due to the
unconsidered effects of galaxy clustering.}
\label{vmax_var}
\end{figure}

In order to arrive at an apparent magnitude limit for the CADIS and
A1689 fields, values of $\langle V/V_{\rm max}\rangle$ were
calculated for different
applied limiting magnitudes and plotted as shown in Figure
\ref{vmax_var}. The grey region in this plot corresponds to the
$1\sigma$ errors described by equation (\ref{eq_sigma_v_vmax_av})
which lessen at the fainter limiting magnitudes due to the inclusion
of more objects. Clustering adds extra noise and so these errors are
an underestimate of the true uncertainty (one can show that the
uncertainty in $V/V_{\rm max}$ increases by the square root of the average
number of objects per cluster).

Without knowledge of the effects of clustering, Figure \ref{vmax_var}
shows that a limiting magnitude of $m_{\rm B}\leq 24.5$ for CADIS and
$m_{\rm B}\leq 23.6$ for A1689 corresponds to a value of $\langle
V/V_{\rm max}\rangle \simeq 0.5$ and thus completeness.  These
magnitudes are in agreement with the apparent magnitude limits at
which the number counts begin to fall beneath that measured by deeper
surveys (such as the B band observations of Lilly, Cowie \& Gardner
(1991) which extend to a depth of $m_{\rm B}\simeq 26$) and also
correspond to a $10\sigma$ object detection threshold deduced from the
photometry.

\begin{figure}
\vspace{4mm}
\epsfxsize=80mm
{\hfill
\epsfbox{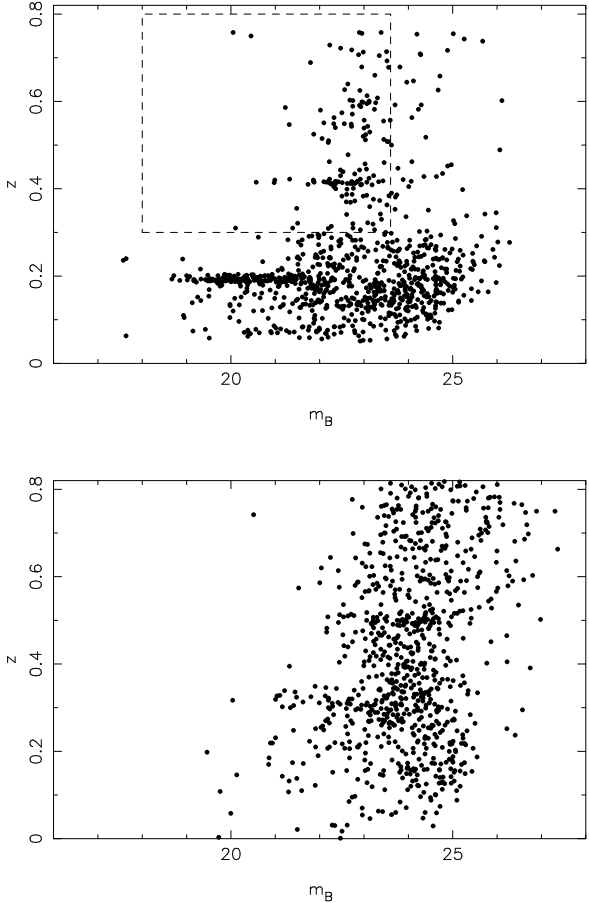}
\hfill}
\epsfverbosetrue
\caption{\small The Abell 1689 (top) and CADIS $16^{\rm h}$ (bottom) field
redshift--apparent B magnitude parameter space. The dashed box
highlights the selection criteria for the 146 A1689 
background galaxies.}
\label{m_z}
\end{figure}

Figure \ref{m_z} shows the distribution of galaxies in the
redshift--apparent magnitude plane in the Abell 1689 field. Abell 1689
itself can clearly been seen at $z=0.18$. Superimposed is the region
of parameter space we use for the mass determination with $0.3<z<0.8$
and $18 \leq m_{\rm B} \leq 23.6$ yielding 146 galaxies.

\subsection{Morphological Type}
\label{sec_morph}

Perhaps the most difficult inconsistency to quantify is that of
variation in galaxy morphological type between samples. It has been
known for some time that elliptical galaxies cluster more strongly
than spirals (Davis \& Geller 1976) and that the elliptical fraction
in clusters is an increasing function of local density (Dressler
1980).  One might therefore expect a sample of galaxies lying behind a
large cluster such as A1689 to contain a higher proportion of
ellipticals than a sample of field galaxies away from a cluster
environment. Since the luminosity function of E/S0 galaxies is thought
to be different from that of spirals (see, for example, Chiba \&
Yoshii 1999 and references therein), comparison of our A1689 sample
with the CADIS offset field sample may be expected to introduce a
bias.

A related issue stems from the fact that the determination of an
object's photometric redshift requires its detection in every filter
belonging to the filter set. Both the CADIS and A1689 filter sets
contain narrow band filters which cause the main restriction on which
objects enter into the photometric analysis. Our $V/V_{\rm max}$ test
in the B-band therefore does not give a true limiting B magnitude but
one which applies only to objects detectable across all filters.  As
long as both samples under comparison are complete according to this
test, the sole consequence of this is that certain morphological types
will be under-represented. With the CADIS filter set of 16 filters
differing from the 9 filters used for the observation of A1689, this
again might be expected to cause inconsistent galaxy types
between fields, biasing results.

Fortunately, our photometric analysis yields morphological type in
addition to redshift. We are therefore able to directly measure the
fraction of galaxies of a given type in both samples and thus test for
biases. We find that in the CADIS sample, the ratio of E/S0:Spiral
galaxies is $12\% \pm 7\%$ and for the A1689 sample, this is $25\% \pm
13\%$ with Poisson errors quoted. This is in reasonable agreement with
the canonical Postman and Geller (1984) E/S0:S fraction for field
galaxies of about 30\%.  Approximately $60\%-70\%$ of both samples are
classified as starburst galaxies. Given the uncertainty in these
fractions, we would argue that as far as we can tell, they are
consistent with each other.  Without a bigger sample of galaxies and
possibly spectroscopically confirmed morphologies we are unable to do
better although as the measurements stand, we would not expect any
serious inconsistencies in morphology which would bias our results.

\section{Mass Determination}
\label{sec_mass_determ}

Taking the CADIS luminosity function as a good estimate of the
intrinsic distribution of A1689 source magnitudes in the range
$0.3\leq z \leq 0.8$, we can calculate $\kappa_\infty$ and $\Sigma$ assuming
a relation between $\kappa$ and $\gamma$. As noted in Section 
\ref{sec_cadis_field}, in the absence of a detection of evolution
of the luminosity function with redshift, we assume a no-evolution model
for the background sources in this field. In general, the occurrence
of evolution is anticipated however we expect its inclusion in
the background model to have only a minor effect on the derived mass.

For the purpose of comparison with other studies we shall quote 
values of $\kappa$ as well as $\kappa_\infty$ and $\Sigma$. As 
$\kappa$ is dependent on the source redshift, this is not a useful 
quantity to quote when the redshift distribution is known. The 
convergence we quote is the redshift averaged quantity defined by
\be
\label{eq_kappa_scale}
\kappa=\frac{\kappa_{\infty}}{N_b}\sum_{i=1}^{N_b} f(z_i)=
\kappa_{\infty} \lgl f_b \rgl
\ee
where $N_b$ is the number of source galaxies. For the field of Abell 1689
we find that $\langle f_b\rangle =0.57$, giving an effective source 
redshift of $z_{\rm eff}=0.45$.

\subsection{Sources of uncertainty}
\label{sec_source_uncert}

Three sources of error on $\kappa_{\infty}$ were
taken into consideration:
\begin{itemize}

\item[1)] The maximum likelihood error obtained from the width of the
likelihood distribution at $\ln {\cal L}_{\rm max}-0.5\Delta \chi^2$, with
$\Delta\chi^2$ the desired confidence level. All object magnitudes and
redshifts were taken as presented directly in the A1689 data while
assuming the Schechter parameters from equation
(\ref{params_ms_alpha_cadis}).  

\item[2)] The uncertainty of the Schechter parameters $M_*$ and $\alpha$
from the likelihood analysis of the CADIS offset field. 

\item[3)] The redshift and magnitude uncertainties of individual objects
in the A1689 data, derived from the photometric analysis. 

\end{itemize}
In Section \ref{sec_sn_calcs} we will show that the
contribution of each source of uncertainty to the overall error
depends on the number of galaxies included in the analysis. Taking all
146 galaxies across the entire field of view, the errors from each
contributor listed above, expressed as a percentage of the total
standard deviation were found to be; 50\% from the maximum likelihood
(essentially the shot noise), 25\% from the uncertainty in $M_*$ and
$\alpha$ and 25\% from the redshift and magnitude error.

The latter two sources of error in the above list were simultaneously
included using the Monte Carlo method. 1000 simulations were carried
out, randomly drawing values of $M_*$ and $\alpha$ from the convolved
likelihood distribution shown in Figure \ref{cadis_likelihood}. For
each realisation, redshifts and absolute magnitudes of objects in the
A1689 field were scattered in exactly the same fashion as before with
the CADIS dataset using their associated photometric errors. The
standard deviation of the scattered values of $\kappa_{\infty}$
produced in this way was then added in quadrature to the uncertainty
of the maximum likelihood error obtained from item one of the list
above to give the overall error.

The magnitude calibration error of $\sigma_{\Delta M}=0.01$ discussed in
Section \ref{sec_photom} was ignored. Inspection of the form of the
Schechter function in equation (\ref{eq_mag_schechter_function}) shows
that a systematic magnitude offset is exactly equivalent to an error
in $M_*$. Clearly, the $1\sigma$ error quoted for $M_*$ in equation 
(\ref{params_ms_alpha_cadis}) completely overwhelms this magnitude 
calibration uncertainty which was therefore deemed insignificant.

Finally, the dependence of our measurement of $\kappa_{\infty}$
on the feature seen in the A1689 redshift distribution at $z\simeq0.4$
was tested. We removed all galaxies
contributing to this peak and re-calculated the results of
the following two sections. Apart from a larger uncertainty due
to the decreased number of objects, we found very little difference
from the results obtained from the full dataset, indicating
that our measurement is not dominated by the concentration of
galaxies at $z\simeq0.4$.

\subsection{The differential radial $\kappa$ profile}
\label{sec_lum_radial_kappa_profile}

Background source objects from the A1689 data set were binned in
concentric annuli about the cluster centre for the calculation of a
radial mass profile.  The relatively small number of
objects contained in the sample however was unfortunately
insufficient to allow computation of a profile
similar in resolution to that of T98.

\begin{figure}
\vspace{4mm}
\epsfxsize=80mm
{\hfill
\epsfbox{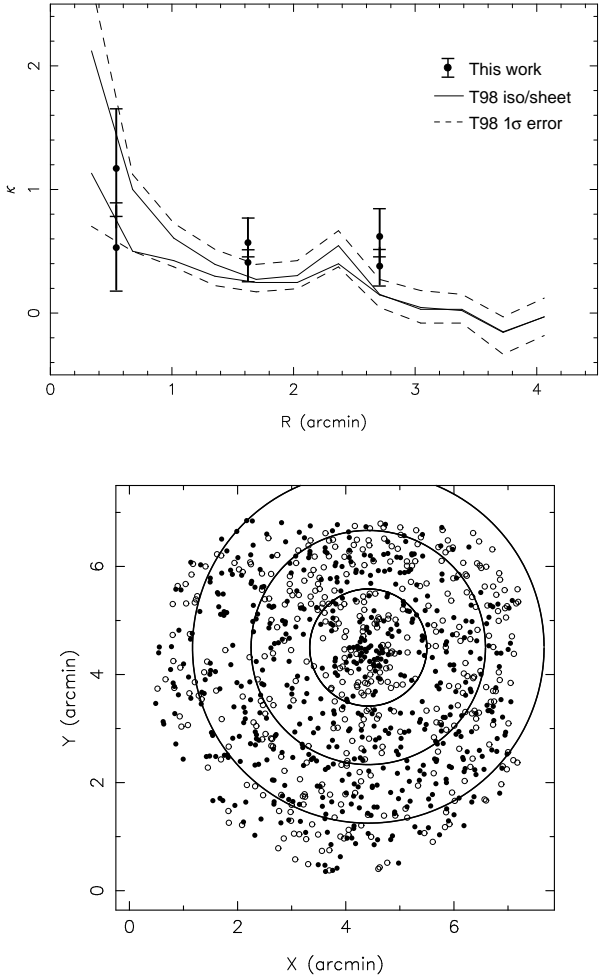}
\hfill}
\epsfverbosetrue
\caption{\small {\em Top} Comparison of radial $\kappa$ profiles for A1689.
Data points show isothermal (lower) and sheet (upper) estimated
$\kappa$ obtained from this work. $1\sigma$ error bars are
plotted. The solid lines indicate the same isothermal/sheet
estimator-bound profile
obtained by T98 using integrated number counts with $1\sigma$ errors
shown by dashed lines. {\em Bottom}: Spatial location of the annular
bins on the A1689 field of view. Open dots are objects selected by
$z>0.2$ and solid dots by $z\leq 0.2$.}
\label{lumfn_kap_profile}
\end{figure}

Apart from the effects of shot noise, this limitation results from the
simple fact that bins which are too narrow do not typically contain a
large enough number of intrinsically bright objects. This has the
effect that the knee of the Schechter function assumed in the
likelihood analysis is poorly constrained. As equation
(\ref{eq_lum_prob}) shows, a large uncertainty in $M_*$ directly results
in a large error on the magnification and hence on $\kappa_{\infty}$.
Experimentation with a range of bin widths quickly showed that in
order to achieve a tolerable precision for $\kappa_{\infty}$, bins had
to be at least $\sim 1.1$ arcmin in width. With the observed field of
view, this gave a limiting number of three bins, illustrated spatially in
the lower half of Figure \ref{lumfn_kap_profile}. In Section \ref{aperture}
we find the average profile within an aperture, which provides a more
robust measurement of $\kappa$.

The top half of Figure \ref{lumfn_kap_profile} shows the 
$\kappa$ data points. These were converted from the
maximum likelihood derived $\kappa_{\infty}$ for each bin using
equation (\ref{eq_kappa_scale}).  The profile of T98 is shown
superimposed for comparison.  Upper points correspond to the sheet
estimator while the lower points are due to the isothermal
estimator. The $1\sigma$ error bars plotted were calculated taking all
three contributions listed in Section \ref{sec_source_uncert} into
account.

Despite relatively large errors, the data points show an amplitude in
good agreement with the profile derived from the number count study of
T98.  These errors seem large in comparison to those of the number
count profile but the number count errors do not take the
systematic uncertainties in background count normalisation, number
count slope or background source redshift into consideration.

It is noticeable that the data points suggest a profile that is perhaps a
little flatter than that derived by T98. It appears
that more mass is detected at larger radii although this is not
particularly significant.

\subsection{Aperture $\kappa$ profile}
\label{aperture}

\begin{figure}
\vspace{4mm}
\epsfxsize=83mm
{\hfill
\epsfbox{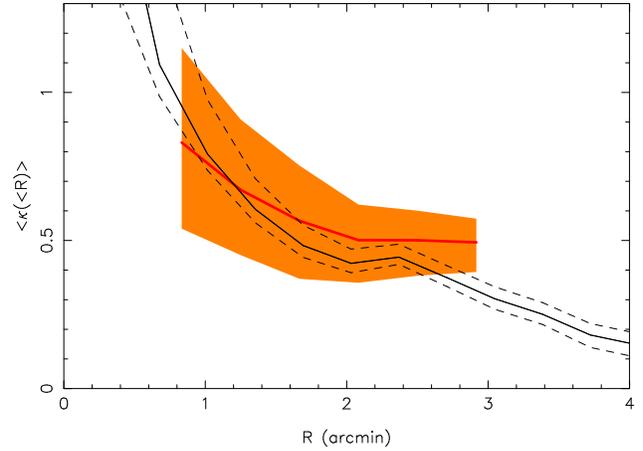}
\hfill}
\epsfverbosetrue
\caption{\small Variation of average surface mass density contained
within a given radius R (thick dark line).  $1\sigma$ errors are shown
by the grey shaded region.  The thin black solid and dashed lines show
the average surface mass density and $1\sigma$ errors (due to
likelihood analysis only) of the parabolic estimated profile of T98.}
\label{aperture_kappa}
\end{figure}

In addition to the radial profile, the variation of average surface
mass density contained within a given radius can be calculated. By
applying the likelihood analysis to the objects contained within an
aperture of varying size, a larger signal to noise can be attained at
larger radii where more objects are encompassed.  With a small
aperture, the same low galaxy count problem is encountered as Figure
\ref{aperture_kappa} shows by the large uncertainty in this vicinity.
In this plot, the parabolic estimator of equation
(\ref{eq_mu_redshift_depen}) is used to obtain $\kappa_{\infty}$ since
as T98 show, this is a good average of the isothermal and sheet
estimators and agrees well their self-consistent axi-symmetric
solution. Application of the axi-symmetric solution is not viable in
this case since we are limited to only 3 bins.

Using equation (\ref{eq_kappa_scale}), $\kappa_\infty$ is again scaled
to $\kappa$.  The grey shaded region in Figure \ref{aperture_kappa}
depicts the $1\sigma$ errors, with the sources of uncertainty from
Section \ref{sec_source_uncert} taken into account.  The thin solid
and dashed black lines show the variation of aperture $\kappa$ and its
error calculated by averaging the parabolic estimator profile
presented in T98. The error does not account for uncertainties arising
from background count normalisation, number count slope or background
source redshift.  The results of T98 were shown to be in good
agreement with the shear analysis of Kaiser (1996) and hence we find a
consistent picture of the mass amplitude and slope from all three
independent methods.

As expected from the results of Section \ref{sec_lum_radial_kappa_profile},
generally more mass than that predicted from the number counts is seen,
especially at large radii. The following section quantifies this
for a comparison with the projected mass result of T98.

\subsection{Projected mass}

From the values of $\kappa_{\infty}$ used to generate the $\kappa$
profile in Section \ref{sec_lum_radial_kappa_profile} and the result
of equation (\ref{eq_kapinf_defn}), the cumulative projected masses in
Table \ref{tab_cum_masses} were calculated. Errors were derived from
propagation of the errors on the binned values of $\kappa_{\infty}$.

\begin{table}
\centering
\begin{tabular}{|c|c|}
\hline
Radius (arcsec) & $M_{2d}(<R)$ \\
\hline
65 & $(0.16\pm0.09)\times 10^{15}h^{-1}{\rm M}_{\odot}$ \\
130 & $(0.48\pm0.16)\times 10^{15}h^{-1}{\rm M}_{\odot}$ \\
195 & $(1.03\pm0.27)\times 10^{15}h^{-1}{\rm M}_{\odot}$ \\
\hline
\end{tabular}
\caption{\small Cumulative projected mass given
by the profile of Section \ref{sec_lum_radial_kappa_profile}.}
\label{tab_cum_masses}
\end{table}

These projected masses are in excellent agreement with those of
T98. At the redshift of the cluster  $1'=0.117\mpcoh$ and hence the second
cumulative mass listed in Table \ref{tab_cum_masses} gives \be
M_{2d}(<0.25\mpcoh)= (0.48\pm0.16)\times 10^{15}\,h^{-1}{\rm
M}_{\odot} \ee which is perfectly consistent with the result from the
number count study. The error here is comparable to the
$30\%$ error quoted for the result of T98 when allowing for all sources
of uncertainty.  The projected mass contained within 195 arcsec is a
little higher than that predicted by T98 although remains arguably 
consistent given the errors involved in each.

\subsection{Effects of sample incompleteness}
\label{sec_incompleteness_effects}

One final uncertainty not taken into consideration so far is that of
sample incompleteness. Changing the limiting apparent B magnitude in
the determination of the CADIS luminosity function directly affects
the fitted values of $M_*$, $\alpha$ and hence the maximum likelihood
$\kappa_{\infty}$.  Similarly, differing numbers of objects
included in the A1689 sample from variations in its limiting magnitude
also has an influence on $\kappa_{\infty}$.

Table \ref{tab_cadis_kapi_var} quantifies this effect for the CADIS
objects. It can be seen that increasing the faint limit $m_{\rm max}$
(ie. including fainter objects) has little effect on $\kappa_{\infty}$
until the limit $m_{\rm max}\simeq 24.5$ is reached.  Beyond this limit,
$\kappa_{\infty}$ starts to fall. Two inferences can therefore be
made. Firstly, this suggests that the magnitude limit in Section
\ref{sec_completeness} from the $V/V_{\rm max}$ test, being
consistent with the limit here, was correctly chosen. Secondly,
$\kappa_{\infty}$ is relatively insensitive to the choice of $m_{\rm max}$
if the sample is complete (and not smaller than the limit at which
shot noise starts to take effect).

\begin{table}
\vspace{2mm}
\centering
\begin{tabular}{|c|cc|ccc|}
\hline
$m_{\rm max}$ & $M_*$ & $\alpha$ & $\kappa_{\infty}({\rm iso})$ &
$\kappa_{\infty}({\rm para})$ & $\kappa_{\infty}({\rm sheet})$ \\
\hline
25.5 & $-19.06$ & $-0.80$ & $0.61^{+0.03}_{-0.04}$ & $0.69^{+0.04}_{-0.06}$ &
$0.76^{+0.07}_{-0.08}$ \\
25.0 & $-19.25$ & $-1.10$ & $0.65^{+0.04}_{-0.04}$ & $0.77^{+0.06}_{-0.06}$ & 
$0.84^{+0.08}_{-0.09}$ \\
24.5 & $-19.43$ & $-1.45$ & $0.70^{+0.06}_{-0.04}$ & $0.85^{+0.08}_{-0.08}$ &
$0.96^{+0.10}_{-0.10}$ \\
24.0 & $-19.98$ & $-1.87$ & $0.74^{+0.03}_{-0.04}$ & $0.91^{+0.13}_{-0.12}$ &
$1.08^{+0.16}_{-0.17}$ \\
23.5 & $-19.24$ & $-1.53$ & $0.75^{+0.04}_{-0.04}$ & $0.90^{+0.06}_{-0.07}$ &
$1.10^{+0.10}_{-0.09}$ \\
\hline
\end{tabular}
\caption{Variation of limiting apparent B magnitude $m_{\rm max}$ of the
CADIS field and its effect on the Schechter parameters and the
resulting value of $\kappa_{\infty}$. Values of $M_*$ assume
$h=1$. The apparent magnitude limit of
$b=23.6$ was assumed for the A1689 data in calculating the maximum
likelihood $\kappa_{\infty}$. Errors here are taken only from the
width of the likelihood curves.}
\label{tab_cadis_kapi_var}
\end{table}

The effect of varying the magnitude limit of the A1689 sample is
quantified in Table \ref{tab_a1689_kapi_var}. A clear trend is also
seen here; as its limiting faint magnitude $m_{\rm max}$ is reduced,
$\kappa_{\infty}$ falls. Assuming linearity, a rough estimate of the
uncertainty of $\kappa_{\infty}$ given the uncertainty of the sample
magnitude limit is given by: 
\be
\Delta\kappa_{\infty}=\left\{\begin{array}{ll} \sim 0.1\Delta m_{\rm max}
& {\rm isothermal} \\ \sim 0.2\Delta m_{\rm max} & {\rm parabolic} \\ \sim
0.4\Delta m_{\rm max} & {\rm sheet} \\
\end{array}\right.
\ee

\begin{table}
\centering
\begin{tabular}{|c|ccc|}
\hline
$m_{\rm max}$ & $\kappa_{\infty}({\rm iso})$ & $\kappa_{\infty}({\rm para})$ &
$\kappa_{\infty}({\rm sheet})$ \\
\hline
24.5 & $0.77^{+0.03}_{-0.03}$ & $1.03^{+0.11}_{-0.11}$ &
$1.30^{+0.15}_{-0.13}$ \\
24.0 & $0.76^{+0.04}_{-0.04}$ & $0.94^{+0.10}_{-0.10}$ &
$1.12^{+0.13}_{-0.12}$ \\
23.5 & $0.69^{+0.06}_{-0.07}$ & $0.79^{+0.10}_{-0.11}$ &
$0.92^{+0.13}_{-0.12}$ \\
\hline
\end{tabular}
\caption{Variation of the maximum likelihood determined $\kappa_{\infty}$
with limiting apparent B magnitude $m_{\rm max}$ of the A1689 data. The
Schechter parameters of Section \ref{sec_cadis_schechter} were assumed
in the likelihood analysis.}
\label{tab_a1689_kapi_var}
\end{table}

Referring to Figure \ref{vmax_var}, a suitable uncertainty in
$m_{\rm max}$ of the A1689 sample of say $\pm0.2$ magnitudes might be
argued. If this were the case, the projected masses of the previous
section calculated with the parabolic estimator would have a further
error of $\sim 5\%$ which is negligibly small.

\section{Signal-to-Noise Predictions}
\label{sec_sn_calcs}

Including all possible contributions of uncertainty in the calculation
of mass, the previous section showed that even with relatively few
galaxies, a significant cluster mass profile can be detected.  One can
make predictions of the sensitivity of the method with differing input
parameters potentially obtained by future measurements.  This exercise
also serves as an optimisation study, enabling identification of
quantities requiring more careful measurement and those which play an
insignificant part.

The most convenient means of carrying out this investigation is by
application of the reconstruction method to simulated galaxy
catalogues. Catalogues were therefore constructed by randomly sampling
absolute magnitudes from the Schechter function fitted to the CADIS
offset field in Section \ref{sec_cadis_schechter}. Redshifts were
assigned to each magnitude by randomly sampling the distribution
parameterised by T98 (their equation 22) from the Canada France
Redshift Survey (Lilly et al. 1995). A range of catalogues were
produced, varying by the number of objects they contained and their
distribution of galaxy redshift errors modeled from the A1689 data.

\begin{figure}
\vspace{4mm}
\epsfxsize=80mm
{\hfill
\epsfbox{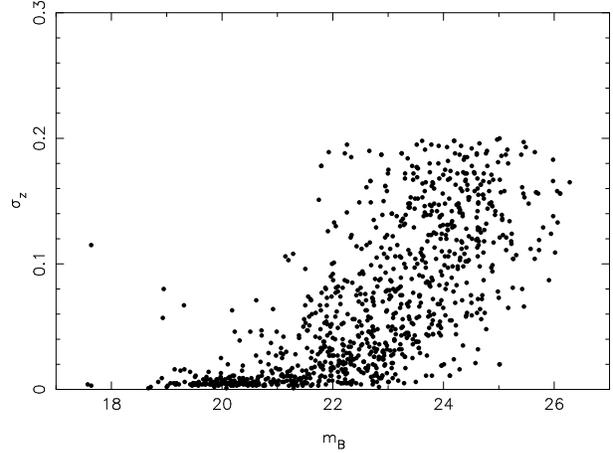}
\hfill}
\epsfverbosetrue
\vspace{-5mm}
\caption{\small Correlation of photometric redshift error with apparent
B-band magnitude for the A1689 data. No significant
correlation between $\sigma_z$ and $z$ exists.}
\label{sz_vs_m}
\end{figure}

Figure \ref{sz_vs_m} shows how the distribution of photometric
redshift error, $\sigma_z$, correlates with the A1689 B-band apparent
magnitude. No significant correlation between $\sigma_z$ and redshift
was found. Catalogue objects were thus randomly assigned redshift
errors in accordance with their apparent magnitude, given by the
correlated distribution in Figure \ref{sz_vs_m}. Different catalogues
were generated from different scalings of this distribution along the
$\sigma_z$ axis.

Each catalogue was then lensed with a sheet mass characterised by
$\kappa_{\infty}=1$ before applying the reconstruction. 1000 Monte Carlo
realisations were performed for each reconstruction, scattering object
redshifts according to their assigned errors in the same manner as in
the reconstruction of A1689. Furthermore, to model the uncertainty
associated with the offset field, assumed values of the Schechter
parameters $M_*$ and $\alpha$ were once again subject to Monte Carlo
realisations. All catalogues were reconstructed assuming sets of
Schechter parameters drawn from a range of scaled versions of the
distribution shown in Figure \ref{cadis_likelihood}.

The resulting scatter measured in the reconstructed value of
$\kappa_{\infty}$ for each catalogue and assumed $\alpha$-$M_*$ scaling
was combined with the average maximum likelihood error across all
realisations of that catalogue to give an overall error. This total
error was found to be well described by,
\be
\label{eq_error_fits}
\sigma_{\kappa_{\infty}}^2 = \frac{1+(2\sigma_z)^2}{n} + 
(0.12\sigma_{\scriptscriptstyle M_*})^2 + (0.37\sigma_{\alpha})^2
-0.18\sigma_{\alpha{\scriptscriptstyle M_*}}
\ee
where $n$ is the number of galaxies, $\sigma_z$ is the sample average
redshift error and $\sigma_{\scriptscriptstyle M_*}$ and
$\sigma_{\alpha}$ are the projected errors on $M_*$ and $\alpha$
respectively as quoted in equation (\ref{params_ms_alpha_cadis}).
The quantity $\sigma_{\alpha{\scriptscriptstyle M_*}}$ is the
covariance of $\alpha$ and $M_*$ defined by
\be
\sigma_{\alpha{\scriptscriptstyle M_*}}=
\frac{\int{\cal L}(M_*,\alpha)(M_*-\langle M_*\rangle)
(\alpha-\langle\alpha\rangle)\,\,\rmd M_* \rmd \alpha}
{\int{\cal L}(M_*,\alpha)\,\,\rmd M_* \rmd \alpha}
\ee
where the likelihood distribution ${\cal L}$ is given by equation
(\ref{eq_schechter_likelihood}). We find that 
$\sigma_{\alpha{\scriptscriptstyle M_*}}=0.039$ for the CADIS
offset field. Equation (\ref{eq_error_fits})
is valid for $n\geq20$ and $0.0 \geq \sigma_z \geq 0.3$.

Equation (\ref{eq_error_fits}) shows that when the number of objects
is low, shot noise dominates. With $n\simeq200$ however, uncertainties
from the calibration of the offset field start to become dominant.
The factor of 2 in the photometric redshift error term stems
from the fact that redshift errors also
translate directly into absolute magnitude errors through equation
(\ref{eq_z_mag_error_prop}). Another discrepancy arises when comparing
this redshift error with the redshift error contribution of 25\%
claimed for the A1689 data in Section \ref{sec_source_uncert}. This is
accounted for by the fact that K-corrections were present in the A1689 
data whereas in the simulated catalogues there were not. 
Equation (\ref{eq_z_mag_error_prop}) quantifies the increase in
magnitude error with the inclusion of K-corrections. This translates to
an approximate increase of $20\%$ in the overall error with
an average K-correction of $-1.0$ for the A1689 data.

Emphasis should be placed on the criteria for which equation
(\ref{eq_error_fits}) is valid. The predicted overall error rises
dramatically when fewer than $\simeq 20$ objects are included in the
analysis. Simulations with 15 objects resulted in maximum likelihood
errors rising to beyond twice that predicted by simple shot
noise. This stems mainly from the effect mentioned in Section
\ref{sec_lum_radial_kappa_profile}, namely
the failure of the likelihood method when the knee of the Schechter
function is poorly constrained.

The most immediate improvement to a multi-colour study such as this
would therefore be to increase galaxy numbers. As previously noted,
only when bins contain $\simeq 200$ objects do offset field
uncertainties become important. Observing in broader filters is one
way to combat the limit presented by galaxy numbers. Section
\ref{sec_gal_cat} noted how the 3000 galaxies detected in the I band
image were instantly reduced to 1000 by the shallow depth limit placed
by the narrow 466/8 filter, even though both were observed to similar
integration times. Using broader filters will also inevitably give
rise to less accurate photometric redshifts. However as the analysis
of this section has shown, one can afford to sacrifice redshift
accuracy quite considerably before its contribution becomes comparable
to that of shot noise.

Deeper observations provide another obvious means of increasing the
number of galaxies.  The error predictions above indicate that the
expected increase in galaxy numbers using an 8 metre class telescope
with the same exposure times as those in this work should reduce shot
noise by a factor of $\sim 3$. Since deeper observations would also
reduce redshift and offset field calibration uncertainties to
negligible levels, the only source of error would be shot noise. In
this case, the signal to noise for $\kappa_{\infty}$ from equation
(\ref{eq_error_fits}) becomes simply
$\kappa_{\infty}\sqrt{n}$ and hence our mass estimate for A1689 could
be quoted with a $9\sigma$ certainty.

\section{Summary}

Photometric redshifts and magnitudes have been determined for objects
in the field of Abell 1689 from multi-waveband observations. This has
allowed calculation of the luminosity function of source galaxies
lying behind the cluster. Comparison of this with the luminosity
function obtained from a similar selection of objects in an
unlensed offset field has resulted in the detection of a bias
in the A1689 background object magnitudes attributed to lens
magnification by the cluster.

To ensure that systematic biases do not affect our results, we have
given careful consideration to the consistency between both the A1689
dataset and the CADIS offset field dataset. We find the distribution
of galaxy types within the redshift range $0.3\leq z \leq 0.8$ applied
to both samples to be very similar. This demonstrates that our lower
redshift cutoff is sufficient to prevent objects in the A1689 dataset
from being significantly influenced by the cluster environment.

After allowing for sources of uncertainty due to redshift and
magnitude error, offset field calibration error and likelihood error
(including shot noise), a significant radial mass profile for A1689
has been calculated.
We predict a projected mass interior to $0.25\mpcoh$ of
\be
M_{2d}(<0.25\mpcoh)=(0.48\pm0.16)\times 10^{15}\,
h^{-1}{\rm M}_{\odot}
\ee
in excellent agreement with the number count analysis of T98 
and the shear results of Tyson \& Fischer (1995) and Kaiser (1995).

We can compare the efficiency of the method presented
in this paper in terms of telescope time and signal-to-noise with the
number count method used by T98. The $5.5\sigma$ result quoted by T98
does not include uncertainty due to their background count
normalisation, number count slope or background source redshift
distribution. Adding these errors to their result gives an estimated
signal-to-noise of $3\sigma$, the same as this work.  Regarding
telescope time, we define a `total information content' for each study
as the product of telescope collecting area and total integration
time. T98 observed 6000s in each of the V and I bands with the NTT
3.6m. Comparing with the CA 3.5m telescope and 12 hours integration
time used in this study shows that we have amassed a total information 
content of approximately 3 times that required for the T98 result.

Despite the extra time penalty induced by our method, we note that
deeper observations, especially in the narrow band filters used, would
increase the signal-to-noise of our result significantly since we are
dominated by shot noise. Our signal-to-noise analyses in Section
\ref{sec_sn_calcs} showed that a $9\sigma$
detection of mass would be possible using an 8 metre class telescope,
equivalent to an increase in integration time by a factor of 5.
This is in contrast to T98 whose main source of
uncertainty comes from the unknown source redshift distribution.
Shot-noise makes a negligible contribution to their error to the
extent that increasing their total integration time by a factor of
three to match the total information content of this work would still
result in a signal-to-noise of $3\sigma$ (Dye 1999).

This paper has been primarily devoted to establishing the viability of
lens mass reconstruction using the luminosity function method. We have
shown that the two main advantages over the number count method employed
by T98 are that use of photometric redshifts have enabled breaking of
the mass/background degeneracy and that the technique is independent
of their clustering if it is assumed that they form an effective random
sampling of luminosity space.

\bigskip
\noindent{\bf ACKNOWLEDGEMENTS} \bib\strut

\noindent
SD thanks PPARC for a studentship and the IfA Edinburgh for funding,
ANT thanks PPARC for a research fellowship and 
EMT thanks the Deutsche Forschungsgemeinschaft for the research fellowship 
supporting his stay in Edinburgh. We also thank
Tom Broadhurst for use of his I-band data and Narciso Benitez who
performed the original data reduction.

\bigskip
\noindent{\bf REFERENCES}
\bib \strut

\bib Athreya R., Mellier Y., van Waerbeke L., Fort B., Pell\'{o} R.
	\& Dantel-Fort M., 1999, submitted to A\&A, astro-ph/9909518

\bib Bertin E. \& Arnouts S., 1996, A\&AS, 117, 393

\bib Broadhurst T.J., Taylor A.N. \& Peacock J.A., 1995, ApJ, 438, 49

\bib Chiba M. \& Yoshii Y., 1999, ApJ, 510, 42

\bib Davis M. \& Geller M.J., 1976, ApJ, 208, 13

\bib Dressler A., 1980, ApJ, 236, 351

\bib Dye S., 1999, PhD Thesis, University of Edinburgh, UK

\bib Ellis R.S., Colless M., Broadhurst T., Heyl J., Glazebrook K., 1996,
	MNRAS, 280, 235

\bib Falco E.E., Gorenstein M.V. \& Shapiro I.I., 1985, ApJ, 437, 56

\bib Fort B., Mellier Y. \& Dantel-Fort M., 1997, A\&A, 321, 353

\bib Francis P.J., Hewett P.C., Foltz C.B., Chaffee F.H., Weymann R.J.,
	Morris S.L., 1991, ApJ, 373, 465

\bib Gray M.E., Ellis R.E., Refregier A., B\'{e}zecourt J., McMahon R.G.,
	Beckett M.G., Mackay C.D. \& Hoenig M.D., 2000,
	submitted to MNRAS, astro-ph/0004161

\bib Gunn J.E. \& Stryker L.L., 1983, ApJS, 52, 121

\bib Kaiser N., 1996, in O. Lahav, E. Terlevich, R.J. Terlevich, eds,
	Proc. Herstmonceux Conf. 36, Gravitational Dynamics,
	Cambridge University Press, p.181

\bib Kinney A.L., Calzetti D., Bohlin R.C., McQuade K., Storchi-Bergmann T.,
	Schmitt H.R., 1996, ApJ, 467, 38

\bib Lilly S.J., Cowie L.L. \& Gardner J.P., 1991, ApJ, 369, 79

\bib Lilly S.J., Tresse L., Hammer F., Crampton D., Le F{\'e}vre O., 1995
	ApJ, 455, 108

\bib Mayen C. \& Soucail G., 2000, submitted to A\&A, astro-ph/0003332

\bib Meisenheimer K., R\"{o}ser H.J., 1996, MPIAPHOT User Manual, 
	MPIA Heidelberg

\bib Meisenheimer K. et al., 1998, in S. D'Oderico, A.Fontana, E. Giallongo,
	eds, ASP Conf. Ser. Vol. 146, The Young Universe: Galaxy Formation 
	and Evolution at Intermediate and High Redshift, p.134

\bib Oke J.B., 1990, AJ, 99, 1621

\bib Parzen E., 1963, Proc. Symp. on Time Series Analysis, p.155,
	ed. I.M. Rosenblatt

\bib Pickles A.J. \& van der Kruit P.C., 1991, A\&AS, 91, 1

\bib Postman M. \& Geller M.J., 1984, ApJ, 281, 95

\bib R\"{o}gnvaldsson \"{O}.E., Greve T.R., Hjorth J., Gudmundsson E.H.,
     	Sigmundsson V.S., Jakobsson P., Jaunsen A.O., Christensen L.L.,
	van Kampen E. \& Taylor A.N., 2000, submitted to MNRAS,
	astro-ph/0009045

\bib Sandage A., Tammann G.A. \& Yahil A., 1979, ApJ, 232, 352

\bib Schmidt M., 1968, ApJ, 151, 393

\bib Taylor A.N., Dye S., Broadhurst T.J., Benitez N., van Kampen E.,
	1998, ApJ, 501, 539

\bib Teague P.F., Carter D. \& Gray P.M., 1990, ApJS, 72, 715

\bib Thommes E.M. et al., 1999, 'CADIS Revised Proposal', MPIA Publication

\bib Tyson J.A. \& Fischer P., 1995, ApJ, 446, L55

\bib van Kampen E., 1998, MNRAS, 301, 389

\bib Wolf C., Meisenheimer K., R\"{o}ser H.J., Beckwith S., Fockenbrock H.,
	Hippelein H., Phleps S., Thommes E., 1999, A\&A, 343, 399

\end{document}